\documentstyle[amsmath,amssymb,graphicx]{article}

\def\tr{{\rm tr} \ }

\begin{document}

\thispagestyle{empty}

\baselineskip15pt

\hfill ITEP/TH-22/09

\bigskip

\begin{center}
{\LARGE \bf Exact 2-point function in Hermitian matrix model }\\
%\vspace{0.3cm}
%\emph{} }\\
\vspace{0.4cm}
{\large A.Morozov and Sh.Shakirov}\\
\vspace{0.4cm}
{\em ITEP, Moscow, Russia\\
MIPT, Dolgoprudny, Russia}\\
\end{center}

\centerline{ABSTRACT}

\bigskip

{\footnotesize
J. Harer and D. Zagier have found a strikingly simple generating function \cite{HZ,HZ2} for exact (all-genera) 1-point correlators in the Gaussian Hermitian matrix model. In this paper we generalize their result to 2-point correlators, using Toda integrability of the model. Remarkably, this exact 2-point correlation function turns out to be an elementary function - arctangent. Relation to the standard 2-point resolvents is pointed out. Some attempts of generalization to $3$-point and higher functions are described. }

\section{Introduction}

In quantum field theory, exact computation of correlation functions in all orders of perturbation theory is rarely possible.
At best, we are able to find a few first terms, and study their properties. Only in low-dimensional and/or topological models, exact correlation functions can be sometime calculated. In this paper, we do such calculation in the Gaussian Hermitian matrix model \cite{AMM}-\cite{UFN3}, where the $m$-point correlators are given by the Gaussian integrals

\begin{align}
C_{i_1 \ldots i_m}(N) = \Big< \tr \phi^{i_1} \ldots \tr \phi^{i_m} \Big> = \int\limits_{N \times N} \ \tr \phi^{i_1} \ldots \tr \phi^{i_m} \ \exp\left( - \dfrac{1}{2} \tr \phi^2 \right) \ d \phi
\end{align}
\smallskip\\
over the space of $N \times N$ Hermitian matrices with flat measure, normalised so that $\Big< 1 \Big> = 1$. Originally designed to study random matrices, Hermitian matrix model is deeply connected to random surfaces \cite{Surfaces}. This is because correlators $C_{i_1 \ldots i_m}(N)$ are polynomials in $N$ with integer coefficients, like

\[
\begin{array}{cccc}
\begin{array}{ll}
C_{2} = N^2 , \ \ \ \ \ C_{4} = 2 N^3 + N, \ \ \ \ \ C_{6} = 5 N^4 + 10 N^2, \ \ \ \ \ C_{1,1} = N, \ \ \ \ \
C_{2,2} = N^4 + 2 N^2, \ \ \ \ldots\\
\\
\end{array}
\end{array}
\]
which count the number of matrix Feynman-t'Hooft graphs (ribbon graphs, fat graphs) of different genus, made of $m$ vertices with $i_1, \ldots, i_m$ legs. Such decomposition of correlators into several parts, related to surfaces of different genera, is usually called genus decomposition. For example, correlator $C_4$ gets contributions from two graphs with topology of a sphere (genus 0), and one graph of torus topology (genus 1):
\begin{figure}[!h]
\begin{center} \includegraphics[width=270pt]{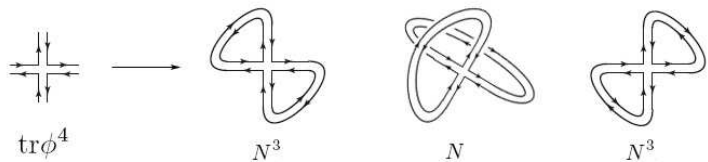}
\caption{Genus decomposition of the correlator $\Big< \tr \phi^4 \Big> = 2N^3 + N$.}
\end{center}
\end{figure}
\smallskip\\
This connection to random surfaces makes matrix models applicable to a wide range of topics in modern physics, including two-dimensional quantum gravity \cite{2dGrav,GKM} and topological string theory \cite{TopString}. Hermitian matrix model has a special place in this spectrum, being one of the simplest and most fundamental models (see \cite{MTheory} for its place in "M-theory of matrix models").

It is therefore interesting and important to calculate Hermitian correlators. As usual in string theory \cite{String}, to represent the answers in a sensible form, it is more convenient to consider a generating function

$$\sum\limits_{i_1 \ldots i_m = 0}^{\infty} C_{i_1 \ldots i_m}(N) \ x_1^{i_1} \ldots x_m^{i_m}$$
\smallskip\\
However, these series do not converge, because the number of fat graphs of arbitrary genus grows too fast. If one wants to cure this problem, one can consider another series, with different normalisation of terms:

$$\sum\limits_{i_1 \ldots i_m = 0}^{\infty} C_{i_1 \ldots i_m}(N) \ \dfrac{x_1^{i_1} \ldots x_m^{i_m} }{n (i_1, \ldots, i_m)} $$
\smallskip\\
Historically, the first example of such a correlation function was constructed by Harer and Zagier \cite{HZ} in the 1-point case. They made the series convergent by special choice of double-factorial weight $n(i)$:

\begin{align}
\sum\limits_{i = 0}^{\infty} C_{2i}(N) \dfrac{x^{2i}}{(2i-1)!!} = \dfrac{1}{2x^2} \left( \left(\dfrac{1 + x^2}{1 - x^2}\right)^N - 1 \right)
\label{HarerZagier}
\end{align}
\smallskip\\
In fact, this choice of weight is quite natural: if one calculates the correlator $C_{2i}(N) = \big< \tr \phi^{2i} \big>$ with the help of Wick theorem, the total number of Wick pairings is exactly $(2i-1)!!$. The result (\ref{HarerZagier}) becomes even simpler, if a generating function with respect to the matrix size $N$ is also calculated \cite{AMM}:

\begin{align}
\sum\limits_{N,i = 0}^{\infty} C_{2i}(N) \dfrac{x^{2i} \lambda^N}{(2i-1)!!}  = \dfrac{\lambda}{1 - \lambda} \dfrac{1}{(1 - \lambda) - (1 + \lambda) x^2}
\label{HarerZagierUni}
\end{align}
\smallskip\\
It should be emphasized that, from the point of view of matrix model theory, this is a highly non-trivial relation: a generating function for 1-point correlators at all genera appears to be \emph{rational}. In this paper we show, that similar relations can be established for the 2-point correlators: a clever choice of the weight $n(i_1, i_2)$ makes the series convergent and, moreover, \emph{elementary} functions. When both $i_1$ and $i_2$ are odd numbers, our result is the following:

\begin{equation}
\addtolength{\fboxsep}{5pt}
\boxed{
\begin{gathered}
\sum\limits_{N,i,j = 0}^{\infty} C_{2i+1, 2j+1}(N) \dfrac{x^{2i+1} y^{2j+1} \lambda^N}{(2i+1)!!(2j+1)!!}  = \dfrac{\lambda}{(\lambda-1)^{3/2}} \dfrac{\arctan\left( \dfrac{xy \sqrt{\lambda-1}}{\sqrt{\lambda - 1 + (\lambda + 1) (x^2 + y^2)}} \right)}{\sqrt{\lambda - 1 + (\lambda + 1) (x^2 + y^2)}}
\end{gathered}
}\label{HarerZagierDuo}
\end{equation}
\smallskip\\
A generating function for correlators with even $i_1$ and $i_2$ is similar, though a little more lengthy, see (\ref{HZ2Even}) below. As one can see, the 2-point generalization of the Harer-Zagier 1-point function is still an elementary function. It is an open question, whether generalized Harer-Zagier correlation functions are always elementary. Affirmative answer to this question would greatly increase our understanding of the model.

It would be certainly interesting to generalize (\ref{HarerZagierUni}) and (\ref{HarerZagierDuo}) to non-Gaussian matrix models, i.e, to Hermitian integrals with non-Gaussian weight. Especially interesting would be generalization to non-Gaussian models with multi-cut support \cite{Multicut1, Multicut2}, which were recently related to supersymmetric gauge theories \cite{Supersymmetry}.

Another interesting direction of generalization are exact correlation functions in the presence of external field (matrix) $\Psi$, see s.2 of \cite{WOperators}. We do not include this topic into the present paper, it will be discussed elsewhere \cite{ExternalField}. The 1-point external-field correlation function turns out to be a simple deformation of the original answer (\ref{HarerZagier}):

\begin{align} \dfrac{1}{2x^2} \left\{ \det _{N \times N} \Big(\dfrac{1 + x^2 + \Psi}{1 - x^2 + \Psi}\Big) - 1 \right\} \end{align}
\smallskip\\
It is again interesting, that no special functions arise even in the external field case. We conclude, that non-standard correlation functions, especially of Harer-Zagier type (with double-factorial weights) appear to be simpler than the standard resolvents (with unity weights) and often can be calculated exactly.

\section{Correlators and integrability}

\subsection{Generalities}
Before doing any actual calculations, let us briefly review the relevant properties of Hermitian matrix model. Partition function of the Hermitian matrix model depends on infinitely many variables $t_k$ known either as coupling constants or as time-variables. Partition function is a formal series in these variables

$$ Z_N (t_1, t_2, \ldots) = \int\limits_{N \times N} \ \exp\left( - \dfrac{1}{2} \tr \phi^2 + \sum\limits_{k} t_k \tr \phi^k \right) d \phi \ = \ 1 + C_i(N) t_i + \dfrac{1}{2!} C_{ij}(N) t_i t_j + \dfrac{1}{3!} C_{ijk}(N) t_i t_j t_k + \ldots $$
\smallskip\\
where coefficients $C_{i_1 \ldots i_m}$ are called $m$-point correlators. Instead of the full correlators (coefficients of $Z_N$) one can consider connected correlators $K_{i_1 \ldots i_m}$, coefficients of the free energy $F_N = \log Z_{N}$:

$$ F_N (t_1, t_2, \ldots) = \log Z_N(t_1, t_2, \ldots) = K_i(N) t_i + \dfrac{1}{2!} K_{ij}(N) t_i t_j + \dfrac{1}{3!} K_{ijk}(N) t_i t_j t_k + \ldots $$
\smallskip\\
In terms of Feynman-t'Hooft diagrams (fat graphs) full correlators count all diagrams, while connected correlators count connected diagrams. This relation between the partition function and its logarithm is not specific for this model, it is a universal property of quantum field theory. At this point it is worth to mention a more traditional notation for correlators, coming from statistical physics:

\begin{align}
C_{i_1 \ldots i_m} = \Big< \tr \phi^{i_1} \ldots \tr \phi^{i_m} \Big>
\end{align}

\begin{align}
K_{i_1 \ldots i_m} = \Big<\Big< \tr \phi^{i_1} \ldots \tr \phi^{i_m} \Big>\Big>
\end{align}
\smallskip\\
As a consequence of relation $F_N = \log Z_N$, the full and connected correlators are related by

\begin{align}
K_{i} \ = \ & C_{i} \\
\nonumber & \\
K_{ij} \ = \ & C_{ij} - C_{i} C_{j} \\
\nonumber & \\
K_{ijk} \ = \ & C_{ijk} - C_{ij} C_{k} - C_{ik} C_{j} - C_{jk} C_{i} + 2 C_{i} C_{j} C_{k}
\end{align}
\smallskip\\
and so on. Because of reflection symmetry of the action $\tr \phi^2$, correlators $C_{i_1 \ldots i_m}$ or $K_{i_1 \ldots i_m}$ are non-vanishing only if $i_1 + \ldots + i_m$ is even. Note, that for $N = 1$ full correlators are quite simple:

\begin{align}C_{i_1 \ldots i_m}(1) = \int\limits_{-\infty}^{+\infty} d \phi \ \phi^{\Sigma i} \ e^{-\phi^2/2} = \left\{ \begin{array}{cc} (\Sigma i - 1)!!, \ \ \Sigma i = \mbox{ even} \\ \\ 0, \ \ \Sigma i = \mbox{ odd} \end{array} \right. \label{N=1}\end{align}
\smallskip\\
where $\Sigma i = i_1 + \ldots + i_m$ is the sum of indices. To avoid confusion, we emphasize once again that in this paper we use normalised Gaussian integrals, i.e, the average of unity is unity:

\begin{align*}\int\limits_{-\infty}^{+\infty} d \phi \ e^{-\phi^2/2} = 1\end{align*}
\smallskip\\
Note also, that there is no difference between 1-point full and connected correlators: $K_i = C_i$. The same is true for 2-point connected correlators with both odd indices: $K_{2i+1, 2j+1} = C_{2i+1, 2j+1}$, as a corollary of above identities and vanishing of 1-point correlators with odd indices, $C_{2i + 1} = 0$. Generally, connected correlators are somewhat simpler and we consider only them from now on.

\subsection{Virasoro constraints}

Usually evaluation of correlators in matrix models
is done with the help of the loop equations, also known
as Ward identities or Virasoro constraints \cite{Virasoro}.
In terms of the partition function, they can be written as

\begin{align} \dfrac{\partial}{\partial t_{b}} Z_N = \sum\limits_{a = 0}^{\infty} a t_{a} \dfrac{\partial Z_N}{\partial t_{a + b - 2}} + \sum\limits_{i + j = b - 2} \dfrac{\partial^2 Z_N}{\partial t_{i} \partial t_{j}}, \ \ \ b > 0 \label{VirasoroZ} \end{align}
\smallskip\\
In terms of free energy $F_N = \log Z_N$, they take form

\begin{align} \dfrac{\partial}{\partial t_{b}} F_N = \sum\limits_{a = 0}^{\infty} a t_{a} \dfrac{\partial F_N}{\partial t_{a + b - 2}} + \sum\limits_{i + j = b - 2} \dfrac{\partial^2 F_N}{\partial t_{i} \partial t_{j}} + \sum\limits_{i + j = b - 2} \dfrac{\partial F_N}{\partial t_{i}} \dfrac{\partial F_N}{\partial t_{j}}, \ \ \ b > 0 \label{VirasoroF}\end{align}
\smallskip\\
The technique based on loop equations \cite{Loopeqs,Virasoro} allows to calculate the correlation functions for arbitrary genus. However, as also mentioned in ref.\cite{AMM}, it turns out that \textit{all-genera} correlation functions, at least in the Gaussian phase, are much simpler
deduced by another method, making explicit use of \textit{integrability} of the model. A drawback of this method is that it is more difficult
to generalize to non-Gaussian phases than the loop equation approach, but in the present paper we are only interested in Gaussian correlators.

\subsection{Integrability}
Our main point in this paper is that correlators in the Hermitian matrix model are constrained by integrable differential equations \cite{UFN3}. In principle, this should allow to calculate exactly all the quantities of interest. In the case of Gaussian Hermitian matrix model, the relevant equations are Toda equations \cite{Toda}, which can be written in terms of the partition function:

\begin{align}
\dfrac{Z_{N+1} Z_{N-1}}{Z_{N}^2} = \dfrac{1}{N} \dfrac{\partial^2}{\partial t_1^2} \log Z_{N}
\label{TodaZ}
\end{align}
\smallskip\\
Equivalently, these equations can be expressed in terms of free energy:

\begin{align}
F_{N+1} - 2 F_{N} + F_{N - 1} = \log\left( \dfrac{1}{N} \dfrac{\partial^2}{\partial t_1^2} F_{N} \right)
\label{TodaF}
\end{align}
\smallskip\\
To calculate the correlation functions, we need to rewrite the Toda equations as relations between correlators. Differentiating eq. (\ref{TodaF}) by $t_{i_1} \ldots t_{i_m}$ and using the connection between partition function and correlators

\begin{align}
C_{i_1 \ldots i_m}(N) = \dfrac{\partial^m}{\partial t_{i_1} \ldots \partial t_{i_m}} Z_N \ \Big|_{t=0}\\
\nonumber\\
K_{i_1 \ldots i_m}(N) = \dfrac{\partial^m}{\partial t_{i_1} \ldots \partial t_{i_m}} F_N \ \Big|_{t=0}
\end{align}
\smallskip\\
one obtains the desired relations. Let us derive them explicitly, in the case of one- and two-point correlators.

\subsection{1-point correlators} In the 1-point case, differentiating eq. (\ref{TodaF}) by $t_i$ we obtain

\begin{align}
K_{i}(N+1) - 2 K_i(N) + K_{i}(N-1) = \dfrac{1}{N} K_{i,1,1}(N)
\label{Discrete1point}
\end{align}
\smallskip\\
where we have used a simple identity $K_{1,1} = C_{11} = N$. Another identity we will need is the following:

\begin{align}
K_{i,1,1}(N) = i(i-1) K_{i-2}(N)
\label{TwoUnities1point}
\end{align}
\smallskip\\
This identity is a corollary Virasoro constraints (\ref{VirasoroF}). Indeed, the Virasoro constraint for $b = 1$ implies

\begin{align} \dfrac{\partial}{\partial t_{1}} F_N = \sum\limits_{a = 0}^{\infty} a t_{a} \dfrac{\partial F_N}{\partial t_{a - 1}} \end{align}
\smallskip\\
Differentiating by $t_1$ and using the Virasoro constraint once again, we get

\begin{align} \dfrac{\partial^2}{\partial t_{1}^2} F_N = N + \sum\limits_{p,q = 0}^{\infty} pq t_{p} t_{q} \dfrac{\partial^2 F_N}{\partial t_{p - 1} \partial t_{q - 1}} + \sum\limits_{a = 0}^{\infty} a (a-1) t_{a} \dfrac{\partial F_N}{\partial t_{a - 2}} \end{align}
\smallskip\\
which justifies (\ref{TwoUnities1point}). Substituting (\ref{TwoUnities1point}) into (\ref{Discrete1point}), we obtain

\begin{equation}
\addtolength{\fboxsep}{5pt}
\boxed{
\begin{gathered}
K_{i}(N+1) - 2 K_i(N) + K_{i}(N-1) = \dfrac{i(i-1)}{N} K_{i-2}(N)
\end{gathered}
}\label{xDiscrete1point}
\end{equation}
\smallskip\\
This system of recursive relations uniquely determines all the one-point correlators $K_i(N)$ by reducing them to the $N = 1$ correlators $K_i(1)$, which are already quite trivial:

\begin{align}K_{2i}(1) = C_{2i}(1) = \int\limits_{-\infty}^{+\infty} d \phi \ \phi^{2i} \ e^{-\phi^2/2} = (2i-1)!!\end{align}
\smallskip\\
Interestingly, 1-point correlators satisfy additional recursive relations, different from (\ref{xDiscrete1point}):

\begin{equation}
\addtolength{\fboxsep}{5pt}
\boxed{
\begin{gathered}
K_{i}(N+1) - K_{i}(N-1) = \dfrac{i + 2}{N} K_{i}(N)
\end{gathered}
}\label{Strange}
\end{equation}
\smallskip\\
These relations are similar to (\ref{xDiscrete1point}), but look slightly simpler. However, they do not follow from Toda equations and their generalization to 2-point and higher-point correlators so far remains unclear. We do not use them in this paper, but it is important to mention that they exist.

\subsection{2-point correlators}

We now do the analogous calculation in the two-point case. Differentiating eq. (\ref{TodaF}) by $t_i$ and $t_j$, we obtain

\begin{align}
K_{ij}(N+1) - 2 K_{ij}(N) + K_{ij}(N-1) = \dfrac{1}{N} K_{i,j,1,1}(N) - \dfrac{1}{N^2} K_{i,1,1}(N) K_{j,1,1}(N)
\label{Discrete2point}
\end{align}
\smallskip\\
As a corollary of Virasoro constraints,

\begin{align}
K_{i,j,1,1}(N) = i(i-1) K_{i-2,j}(N) + 2 ij K_{i-1,j-1}(N) + j(j-1) K_{i,j-2}(N)
\label{TwoUnities2point}
\end{align}
\smallskip\\
Substituting (\ref{TwoUnities2point}) into (\ref{Discrete2point}), we obtain recursive relations for the 2-point correlators:

\begin{equation}
\addtolength{\fboxsep}{5pt}
\boxed{
\begin{gathered}
K_{ij}(N+1) - 2 K_{ij}(N) + K_{ij}(N-1) =  - \dfrac{ij(i-1)(j-1)}{N^2} K_{i-2}(N) K_{j-2}(N) + \emph{} \\
\\
\emph{} + \dfrac{1}{N} \Big( i(i-1) K_{i-2,j}(N) + 2 ij K_{i-1,j-1}(N) + j(j-1) K_{i,j-2}(N) \Big)
\end{gathered}
}\label{xDiscrete2point}
\end{equation}
\smallskip\\
In complete analogue with the 1-point case, these recursive relations allow to express arbitrary correlators $K_{ij}(N)$ through the $N = 1$ correlators, which are simple according to (\ref{N=1}). Note however, that there are two essentially different cases: when both indices are even

\begin{align}K_{2i, 2j}(1) = C_{2i, 2j}(1) - C_{2i}(1) C_{2j}(1) = (2i + 2j - 1)!! - (2i - 1)!! (2j - 1)!!\end{align}
\smallskip\\
and when both indices are odd

\begin{align}K_{2i+1, 2j+1}(1) = C_{2i+1, 2j+1}(1) - C_{2i+1}(1) C_{2j+1}(1) = (2i + 2j + 1)!!\end{align}
\smallskip\\
The second "totally odd" case is somewhat simpler, because in this case the quadratic contribution -- the product of two odd 1-point correlators -- vanishes. This is actually one of the reasons, why the totally odd generating function (\ref{HarerZagierDuo}) is simpler, than its totally even counterpart.

\section{Correlation functions}

There are many different ways to solve discrete relations like (\ref{xDiscrete1point}) or (\ref{xDiscrete2point}). For example, one can start solving them iteratively and try to guess a combinatorial formula. As explained in \cite{AMM}, a more elegant way to proceed is to pass from particular correlators (which depend on discrete indices like $i$ or $j$) to generating functions (which depend on continuous variables). In terms of generating functions, the discrete relations like (\ref{xDiscrete1point}) turn into differential equations which often admit simple and elegant solutions.

For the sake of brevity, we call generating functions for correlators simply correlation functions. These objects are also known in literature as (multi-)densities \cite{AMM}. In this paper, we restrict consideration to three types of correlation functions. First of all, the standard correlation functions, with unity weights:

\begin{align} \rho_{N}(x_1, \ldots, x_m) = \Big< \Big< \tr \left( \dfrac{1}{1 - x_1 \phi} \right) \ldots \tr \left( \dfrac{1}{1 - x_m \phi} \right) \Big> \Big> = \sum\limits_{i_1 \ldots i_m = 0}^{\infty} K_{i_1 \ldots i_m}(N) \ x_1^{i_1} \ldots x_m^{i_m} \end{align}
\smallskip\\
These functions are the most commonly used in literature, especially in the context of loop equations approach \cite{Loopeqs}. To avoid misunderstanding, let us note that in the context of loop equations the same functions are usually defined in a slightly different way

\begin{align} W_{N}(x_1, \ldots, x_m) = \Big< \Big< \tr \left( \dfrac{1}{x_1 - \phi} \right) \ldots \tr \left( \dfrac{1}{x_m - \phi} \right) \Big> \Big> = \sum\limits_{i_1 \ldots i_m = 0}^{\infty} K_{i_1 \ldots i_m}(N) \ x_1^{-i_1-1} \ldots x_m^{-i_m-1} \label{Resolv} \end{align}
\smallskip\\
and known as resolvents. We use the terms "standard correlation function" and "resolvent" as synonims here, because it is not a problem to transform one into another: two different definitions are related by

$$W_{N}(x_1, \ldots, x_m) = \dfrac{1}{x_1 \ldots x_m} \rho_{N}\left(\dfrac{1}{x_1}, \ldots, \dfrac{1}{x_m}\right)$$
\smallskip\\
Therefore, it does not make a big difference. In this paper, \pagebreak we prefer to use functions $\rho_{N}$ rather than functions $W_{N}$. Second, we consider the exponential correlation functions, i.e, with factorial weight:

\begin{align} e_{N}(x_1, \ldots, x_m) = \Big< \Big< \tr \left( e^{x_1 \phi} \right) \ldots \tr \left( e^{x_m\phi} \right)  \Big> \Big> = \sum\limits_{i_1 \ldots i_m = 0}^{\infty} K_{i_1 \ldots i_m}(N) \ \dfrac{x_1^{i_1} \ldots x_m^{i_m}}{i_1! \ldots i_m!}\end{align}
\smallskip\\
The third functions are those which play the central role in our paper -- the Harer-Zagier correlation functions:

\begin{align} \varphi_{N}(x_1, \ldots, x_m) = \sum\limits_{i_1 \ldots i_m = 0}^{\infty} K_{i_1 \ldots i_m}(N) \ \dfrac{x_1^{i_1} \ldots x_m^{i_m}}{n(i_1) \ldots n(i_m)}, \ \ \ \ n(i) = \left\{ \begin{array}{lll} (i-1)!!, \ \ \ i = {\rm even} \\ \\ i!!, \ \ \ i = {\rm odd} \\ \end{array}\right. \end{align}
\smallskip\\
Obvoiusly, different choices of weights can be useful under different circumstances. The Harer-Zagier functions are useful just because they provide simple and explicit answers: it is enough to take a look at (\ref{HarerZagierUni}) or (\ref{HarerZagierDuo}). The exponential and standard correlation functions are useful for another reason: it is because Toda equations, rewritten as differential equations on these functions, take the simplest form. This can be seen already for the 1-point equations (\ref{xDiscrete1point}), since the term $i(i-1) K_{i-2}$ suggests two natural choices of weights:

$$\sum\limits_i i(i-1) K_{i-2} \dfrac{z^i}{i!} = x^2 \left( \sum\limits_{i} K_i \dfrac{z^i}{i!} \right) $$
\smallskip\\
and

$$\sum\limits_i i(i-1) K_{i-2} z^{-i-1} = \dfrac{\partial^2}{\partial x^2} \left( \sum\limits_{i} K_i z^{-i-1} \right)$$
\smallskip\\
The first choice corresponds to the exponential correlation function, while the second choice corresponds to the resolvent in its usual form (\ref{Resolv}). For other choices of weights, operator in the right hand side would be more complicated than just $x$ squared or the second derivative. That is why, in a sence, these two choices are singled out by the equation itself.

Of course, the three correlation functions are related by various integral transformations. First, the standard and exponential functions are related just by Laplace transform:

\begin{align} \rho_{N}(x_1, \ldots, x_m) = \int\limits_{0}^{\infty} dy_1 \ldots \int\limits_{0}^{\infty} dy_m \ e_{N}(x_1 y_1, \ldots, x_m y_m) \ e^{- y_1 - \ldots - y_m}  \end{align}
\smallskip\\
Second, the standard and Harer-Zagier functions are related by a certain Gaussian transform:

\begin{align} \rho_{N}(x_1, \ldots, x_m) = \int\limits_{-\infty}^{\infty} (1 + y_1) dy_1 \ldots \int\limits_{-\infty}^{\infty} (1 + y_m) dy_m \ \varphi_{N}(x_1 y_1, \ldots, x_m y_m) \ e^{- y_1^2/2 - \ldots - y_m^2/2} \end{align}
\smallskip\\
Third, the exponential and Harer-Zagier functions are related by a contour-integral transform:

\begin{align} e_{N}(x_1, \ldots, x_m) = \oint (1 + y_1) dy_1 \ldots \oint (1 + y_m) dy_m \ \dfrac{\varphi_{N}(y_1, \ldots, y_m)}{y_1 \ldots y_m} \ \exp\left( \dfrac{x_1^2}{2y_1^2} + \ldots + \dfrac{x_m^2}{2y_m^2} \right)\label{ContourTransform}\end{align}
\smallskip\\
If one is able to find one of these functions -- $e$, $\rho$ or $\varphi$ -- it is not a problem to convert it into another. Therefore, one can freely change the weights in order to simplify the solution. Another important simplification, which was suggested in \cite{AMM} and which we intensively use, is to consider the universal generating functions, i.e, generating functions w.r.t $N$ with parameter $\lambda$:

\begin{align}
\rho(\lambda; x_1, \ldots, x_m) = \sum\limits_{N = 0}^{\infty} \ \lambda^N \ \rho_{N}(x_1, \ldots, x_m)\\
\nonumber \\
e(\lambda; x_1, \ldots, x_m) = \sum\limits_{N = 0}^{\infty} \ \lambda^N \ e_{N}(x_1, \ldots, x_m)\\
\nonumber \\
\varphi(\lambda; x_1, \ldots, x_m) = \sum\limits_{N = 0}^{\infty} \ \lambda^N \ \varphi_{N}(x_1, \ldots, x_m)
\end{align}
\smallskip\\
As we will see below, transition to universal functions greatly simplifies both the equations and the answer. We are now going to rewrite eqs. (\ref{xDiscrete1point}) and (\ref{xDiscrete2point}) as differential equations on correlation functions and solve them directly. The solution in the 1-point case is the well-known Harer-Zagier function (\ref{HarerZagierUni}). In the 2-point case, its generalization is obtained.

\section{Harer-Zagier correlation functions}

\subsection{1-point function}

Rewritten in terms of the generating function for 1-point correlators

$$\varphi_N(x) = \sum\limits_{i = 0}^{\infty} K_{2i}(N) \dfrac{x^{2i}}{(2i-1)!!}$$
\smallskip\\
eq. (\ref{xDiscrete1point}) becomes a differential equation of first order:

\begin{align}
\varphi_{N+1}(x) - 2 \varphi_{N}(x) + \varphi_{N-1}(x) = \dfrac{1}{N} x \dfrac{\partial}{\partial x} \Big( x^2 \varphi_{N}(x) \Big)
\label{x1point}
\end{align}
\smallskip\\
Passing to generating functions with respect to $N$, we obtain

\begin{align}
\lambda \dfrac{\partial}{\partial \lambda} \left( \dfrac{(1-\lambda)^2}{\lambda} \varphi(\lambda; x) \right) = x \dfrac{\partial}{\partial x} \Big( x^2 \varphi(\lambda; x) \Big)
\label{xx1point}
\end{align}
\smallskip\\
This is a first-order partial differential equation with general solution

\begin{align} \varphi(\lambda; x) = \dfrac{\lambda}{(\lambda - 1)^2 x^2} \ U \left( \dfrac{1}{x^2} + \dfrac{2}{\lambda - 1} \right) \end{align}
\smallskip\\
The function $U(z)$ is determined from the initital condition

\begin{align}
\dfrac{\partial}{\partial \lambda} \varphi(\lambda; x) \Big|_{\lambda = 0} = \varphi_{N = 1}(x) = \sum\limits_{i = 0}^{\infty} \dfrac{x^{2i}}{(2i-1)!!} \int\limits_{-\infty}^{+\infty} d \phi \ \phi^{2i} \ e^{-\phi^2/2} = \sum\limits_{i = 0}^{\infty} x^{2i} = \dfrac{1}{1 - x^2}
\label{initcond1}
\end{align}
\smallskip\\
In other words, the initial condition is merely a consequence of the fact that $K_i(1) = (2i-1)!!$ and the choice of weights. Comparing the general solution with the initial condition, we obtain  $U(z) = 1/(1+z)$ and

\begin{equation}
\addtolength{\fboxsep}{5pt}
\boxed{
\begin{gathered}
\varphi(\lambda; x) = \dfrac{\lambda}{1 - \lambda} \dfrac{1}{(1 - \lambda) - (1 + \lambda) x^2}
\end{gathered}
}\label{HZ1}
\end{equation}
\smallskip\\
In this way the Harer-Zagier correlation function $\varphi(\lambda;x)$ can be found as solution to a linear differential equation of first order in two variables $\lambda$ and $x$. Note, that the universal $\lambda$-dependent correlation function, which contains information for all dimensions $N$, seems to be very similar to the $N = 1$ function: namely, they are related by multiplicative transform

\begin{align}
 x^2 \mapsto \dfrac{1 + \lambda}{1 - \lambda} x^2, \ \ \ \varphi \mapsto \dfrac{\lambda}{(1 - \lambda)^2} \varphi
\label{1pPattern}
\end{align}
\smallskip\\
Interestingly, this property has a literal analogue in the 2-point case, see (\ref{2pPattern}) below.

\subsection{2-point function}

For two-point correlators, the two separate generating functions can be introduced, even and odd:

\begin{align}
\varphi^{+}_N(x,y) = \sum\limits_{i,j = 0}^{\infty} K_{2i,2j}(N) \dfrac{x^{2i} y^{2j}}{(2i-1)!!(2j-1)!!}\\
\nonumber \\
\varphi^{-}_N(x,y) = \sum\limits_{i,j = 0}^{\infty} K_{2i+1,2j+1}(N) \dfrac{x^{2i+1} y^{2j+1}}{(2i+1)!!(2j+1)!!}
\end{align}
\smallskip\\
In terms of these generating functions, eq. (\ref{xDiscrete2point}) becomes a system of two equations:

$$ \varphi^{+}_{N+1}(x,y) - 2 \varphi^{+}_{N}(x,y) + \varphi^{+}_{N-1}(x,y) = $$

\begin{align}
= \dfrac{1}{N} \ \left( x \dfrac{\partial}{\partial x} x^2 + y \dfrac{\partial}{\partial y} y^2 \right) \varphi^{+}_{N}(x,y) + \dfrac{2}{N} \ xy \dfrac{\partial^2}{\partial x \partial y} x y \ \varphi^{-}_{N}(x,y) - \dfrac{1}{N^2} \ \left( xy \dfrac{\partial^2}{\partial x \partial y} x^2 y^2 \right) \varphi_{N}(x) \varphi_{N}(y)
\label{x2point1}
\end{align}
\smallskip\\
and

\begin{align}
\varphi^{-}_{N+1}(x,y) - 2 \varphi^{-}_{N}(x,y) + \varphi^{-}_{N-1}(x,y) = \dfrac{1}{N} \ \left( x^2 \dfrac{\partial}{\partial x} x + y^2 \dfrac{\partial}{\partial y} y \right) \varphi^{-}_{N}(x,y) + \dfrac{2}{N} \ xy \ \varphi^{+}_{N}(x,y)
\label{x2point2}
\end{align}
\smallskip\\
Passing to generating functions with respect to $N$, we obtain

\begin{align}
\left\{
\begin{array}{ll}
\lambda \dfrac{\partial}{\partial \lambda} \left( \dfrac{(1-\lambda)^2}{\lambda} \varphi^{+}(\lambda; x,y) \right) = \left( x \dfrac{\partial}{\partial x} x^2 + y \dfrac{\partial}{\partial y} y^2 \right) \varphi^{+}(\lambda; x,y) + 2 xy \dfrac{\partial^2}{\partial x \partial y} x y \varphi^{-}(\lambda; x,y)  - G(\lambda; x,y) \\
\\
\lambda \dfrac{\partial}{\partial \lambda} \left( \dfrac{(1-\lambda)^2}{\lambda} \varphi^{-}(\lambda; x,y) \right) = \left( x^2 \dfrac{\partial}{\partial x} x + y^2 \dfrac{\partial}{\partial y} y \right) \varphi^{-}(\lambda; x,y) + 2 x y \varphi^{+}(\lambda; x,y) \\
\end{array}\right.
\label{xx2point}
\end{align}
\smallskip\\
or in a matrix form

\[
\left( \begin{array}{cc}
\lambda \ \dfrac{\partial}{\partial \lambda} \ \dfrac{(1-\lambda)^2}{\lambda} - x \dfrac{\partial}{\partial x} x^2 - y \dfrac{\partial}{\partial y} y^2 & - 2 xy \dfrac{\partial^2}{\partial x \partial y} x y \\
\\
- 2 xy & \lambda \ \dfrac{\partial}{\partial \lambda} \ \dfrac{(1-\lambda)^2}{\lambda} - x^2 \dfrac{\partial}{\partial x} x - y^2 \dfrac{\partial}{\partial y} y
\end{array} \right) \ \left( \begin{array}{cc} \varphi^{+} \\ \\ \varphi^{-} \end{array} \right) = \left( \begin{array}{cc} G \\ \\ 0 \end{array} \right)
\]
\smallskip\\
where the free term $G(\lambda; x,y)$ is given by

\begin{align} G(\lambda; x,y) = \sum\limits_{N = 1}^{\infty} \dfrac{\lambda^N}{N} \ \left(  xy \dfrac{\partial^2}{\partial x \partial y} x^2 y^2 \right) \ \varphi_{N}(x) \ \varphi_{N}(y) = \lambda \left( \dfrac{2 x y}{(\lambda - 1)(1 + x^2 y^2) + (\lambda + 1) (x^2 + y^2)} \right)^2 \end{align}
\smallskip\\
This system of two differential equations is more complicated, than in the 1-point case, but it is still linear and can be solved by elementary means. First of all, we need to find the initial conditions, what is done again by comparing to the $N = 1$ case. For $N = 1$, we know the correlators explicitly

$$K_{2i,2j}(1) = C_{2i,2j}(1) - C_{2i}(1) C_{2j}(1) = (2i+2j-1)!! - (2i-1)!! (2j-1)!!$$

$$K_{2i+1,2j+1}(1) = C_{2i+1,2j+1}(1) = (2i+2j+1)!! $$
\smallskip\\
so we can compute the generating functions

$$
\varphi^{-}_{N = 1}(x,y) = \sum\limits_{i,j = 0}^{\infty} \dfrac{(2i + 2j + 1)!!}{(2i+1)!!(2j+1)!!} x^{2i + 1} y^{2j + 1} = \dfrac{1}{\sqrt{1 - x^2 - y^2}} \arctan\left( \dfrac{xy}{\sqrt{1 - x^2 - y^2}} \right)
$$

$$
\varphi^{+}_{N = 1}(x,y) = \sum\limits_{i,j = 0}^{\infty} \dfrac{(2i + 2j - 1)!! - (2i-1)!!(2j-1)!!}{(2i-1)!!(2j-1)!!} x^{2i} y^{2j} = \dfrac{xy}{x^2-y^2} \left( x \dfrac{\partial}{\partial x} - y \dfrac{\partial}{\partial y} \right) \varphi^{-}_{N = 1}(x,y)
$$
\smallskip\\
Therefore, initial conditions for (\ref{xx2point}) are

\begin{align}
\left\{
\begin{array}{ll}
\dfrac{\partial}{\partial \lambda} \varphi^{-}(\lambda; x,y) \Big|_{\lambda = 0} = \varphi^{-}_{N = 1}(x,y) = \dfrac{1}{\sqrt{1 - x^2 - y^2}} \arctan\left( \dfrac{xy}{\sqrt{1 - x^2 - y^2}} \right) \\
\\
\dfrac{\partial}{\partial \lambda} \varphi^{+}(\lambda; x,y) \Big|_{\lambda = 0} = \varphi^{+}_{N = 1}(x,y) = \dfrac{xy}{x^2-y^2} \left( x \dfrac{\partial}{\partial x} - y \dfrac{\partial}{\partial y} \right) \varphi^{-}_{N = 1}(x,y) \\
\end{array}\right.
\label{initcond2}
\end{align}
\smallskip\\
The unique solution of (\ref{xx2point}) with initial conditions (\ref{initcond2}) is the following:

\begin{equation}
\addtolength{\fboxsep}{5pt}
\boxed{
\begin{gathered}
\varphi^{-}(\lambda; x,y) = \dfrac{\lambda}{(\lambda-1)^{3/2}} \dfrac{\arctan\left( \dfrac{xy \sqrt{\lambda-1}}{\sqrt{\lambda - 1 + (\lambda + 1) (x^2 + y^2)}} \right)}{\sqrt{\lambda - 1 + (\lambda + 1) (x^2 + y^2)}}
\end{gathered}
}\label{HZ2Odd}
\end{equation}

\begin{equation}
\addtolength{\fboxsep}{5pt}
\boxed{
\begin{gathered}
\varphi^{+}(\lambda; x,y) \ = \ \dfrac{xy}{x^2-y^2} \left( x \dfrac{\partial}{\partial x} - y \dfrac{\partial}{\partial y} \right) \varphi^{-}(\lambda; x,y) =
\end{gathered}
}\label{HZ2Even}
\end{equation}

\begin{align*}
\nonumber & = \dfrac{\lambda(\lambda + 1) x^2y^2}{(1-\lambda)} \Big( \lambda - 1 + (1 + \lambda) (x^2+ y^2) \Big)^{-1} \Big( \lambda - 1 + (1 + \lambda) (x^2+ y^2) + (\lambda-1) x^2y^2 \Big)^{-1} - \emph{} \\ & \nonumber \\ & \emph{} - \dfrac{\lambda(\lambda+1)xy}{(1 - \lambda)^{3/2}} \Big(\lambda - 1 + (\lambda + 1) (x^2 + y^2)\Big)^{-3/2} \ \arctan\left( \dfrac{xy \sqrt{\lambda-1}}{\sqrt{\lambda - 1 + (\lambda + 1) (x^2 + y^2)}} \right)
\end{align*}
\smallskip\\
It is an elementary exercise to substitute (\ref{HZ2Odd}) and (\ref{HZ2Even}) into (\ref{xx2point}) and check that equations are satisfied. The check of initial conditions (\ref{initcond2}) is equally simple. Just like in the 1-point case, the universal $\lambda$-dependent correlation function is related to the $N = 1$ correlation function by a simple transformation

\begin{align}
 x^2 + y^2 \mapsto \dfrac{1 + \lambda}{1 - \lambda} (x^2 + y^2), \ \ \ xy \mapsto xy, \ \ \ \varphi \mapsto \dfrac{\lambda}{(1 - \lambda)^2} \varphi
\label{2pPattern}
\end{align}
\smallskip\\
This striking relation, valid for both even and odd functions, is clearly a hint for some larger structure, which can be completely revealed only by the study of the 3-point and higher-point cases. Note also, that it is straightforward to extract correlation functions for particular $N$ from the universal $\lambda$-dependent correlation function. However, the formulas become less explicit, i.e, involving an integral:

\begin{align}
 \varphi^{-}_N(x,y) = \int\limits_{0}^{xy} \dfrac{dt}{2(x^2+y^2)} \ \left( \left( \dfrac{1 + t^2 + x^2 + y^2}{1+t^2-x^2-y^2} \right)^N - 1 \right)
\label{U1}
\end{align}

\begin{align}
\varphi^{+}_N(x,y) = \int\limits_{0}^{xy} \dfrac{xy dt}{(x^2+y^2)^2} \ \left( \left( \dfrac{1 + t^2 + x^2 + y^2}{1+t^2-x^2-y^2} \right)^N \dfrac{2N(x^2+y^2)(1+t^2)+(x^2+y^2)^2-(1+t^2)^2}{(1 + t^2 + x^2 + y^2)(1 + t^2 - x^2 - y^2)} + 1 \right)
\label{U2}
\end{align}
\smallskip\\
Formulas (\ref{HZ2Odd}) and (\ref{HZ2Even}) completely describe the exact 2-point correlators: it is enough to write

$$ \dfrac{ \Big<\Big< \tr \phi^{2i+1} \tr \phi^{2j+1}  \Big> \Big> }{(2i+1)!!(2j+1)!!} = \mbox{ coefficient of } x^{2k+1} y^{2m+1} \lambda^N \mbox{ in } \dfrac{\lambda}{(\lambda-1)^{3/2}} \dfrac{\arctan\left( \dfrac{xy \sqrt{\lambda-1}}{\sqrt{\lambda - 1 + (\lambda + 1) (x^2 + y^2)}} \right)}{\sqrt{\lambda - 1 + (\lambda + 1) (x^2 + y^2)}} $$
\smallskip\\
and similarly for even correlators. Generalisation of these formulas to 3-point and higher-point cases is not straightforward, since Toda equations become more complicated and explicit solution becomes increasingly hard to find. Generalization to non-Gaussian (say, Djkgraaf-Vafa \cite{Multicut2}) models is even more obscure, since integrable equations (\ref{TodaZ}) are non-trivially modified in these models.

\section{Exponential correlation functions}

\subsection{Recursive relations}
Direct generalization of (\ref{HZ2Odd}) and (\ref{HZ2Even}) looks problematic. To bypass those difficulties, let us use the freedom in the choice of weight and consider not Harer-Zagier but exponential correlation functions:

$$ e_{N}(x_1, \ldots, x_m) = \Big< \Big< \tr \left( e^{x_1 \phi} \right) \ldots \tr \left( e^{x_m\phi} \right)  \Big> \Big> = \sum\limits_{i_1 \ldots i_m = 0}^{\infty} K_{i_1 \ldots i_m}(N) \ \dfrac{x_1^{i_1} \ldots x_m^{i_m}}{i_1! \ldots i_m!}$$
\smallskip\\
Expressed in terms of these functions, the Toda equations do not contain any differential operators at all:

\[
\begin{array}{ccc}
e_{N+1}(x) + e_{N-1}(x) = 2 e_{N}(x) + \dfrac{x^2}{N} e_{N}(x)\\
\\
\\
e_{N+1}(x,y) + e_{N-1}(x,y) = 2 e_{N}(x,y) + \dfrac{(x+y)^2}{N} e_{N}(x,y) - \dfrac{x^2}{N} \dfrac{y^2}{N} e_N(x) e_N(y) \\
\\
\\
e_{N+1}(x,y,z) + e_{N-1}(x,y,z) = 2 e_{N}(x,y,z) + \dfrac{(x+y+z)^2}{N} e_{N}(x,y,z) - \dfrac{(x+y)^2}{N} \dfrac{z^2}{N} e_N(x,y) e_N(z) - \\
\\
- \dfrac{(x+z)^2}{N} \dfrac{y^2}{N} e_N(x,z) e_N(y)  - \dfrac{(y+z)^2}{N} \dfrac{x^2}{N} e_N(y,z) e_N(x) + 2 \dfrac{x^2}{N} \dfrac{y^2}{N} \dfrac{z^2}{N} e_N(x) e_N(y) e_N(z) \\
\\
\end{array}
\]
\smallskip\\
and so on, generally

\begin{align}
 e_{N+1} + e_{N-1} = c_N e_N + g_N
\label{CombinToda}
\end{align}
\smallskip\\
where

$$
c_N = 2 + \dfrac{2}{N}(x_1 + \ldots + x_m)^2
$$
\smallskip\\
and $g_N(x_1, \ldots, x_m)$ is the function which can be considered as already known -- by recursion:

\[
\begin{array}{ccc}
g_N(x) = 0\\
\\
\\
g_N(x,y) = - \dfrac{x^2}{N} \dfrac{y^2}{N} e_N(x) e_N(y) \\
\\
\\
g_N(x,y,z) = - \dfrac{(x+y)^2}{N} \dfrac{z^2}{N} e_N(x,y) e_N(z) - \dfrac{(x+z)^2}{N} \dfrac{y^2}{N} e_N(x,z) e_N(y) - \\
\\
 - \dfrac{(y+z)^2}{N} \dfrac{x^2}{N} e_N(y,z) e_N(x) + 2 \dfrac{x^2}{N} \dfrac{y^2}{N} \dfrac{z^2}{N} e_N(x) e_N(y) e_N(z) \\
\\
\end{array}
\]
\smallskip\\
We now turn to explicit solution of equations (\ref{CombinToda}). In contrast with the Harer-Zagier case, where solution of Toda equations is a non-trivial procedure, in the exponential case solution has nothing to do with differential equations and is rather simple. We emphasise, that this simplicity is due to the choice of weight.

\subsection{Determinantal solution}
As usual for integrable equations, solution of eq. (\ref{CombinToda}) can be given explicitly in terms of $N \times N$ determinants:

\begin{align}
e_{N} = \det\limits_{N \times N}
\left(
\begin{array}{ccccccccccccccc}
e_1 & -g_1 & g_2 & -g_3 & \ldots \\
\\
1 & c_1 & 1 & 0 & \ldots \\
\\
0 & 1 & c_2 & 1 & \ldots \\
\\
0 & 0 & 1 & c_3 & \ldots \\
\\
\ldots & \ldots & \ldots & \ldots & \ldots \\
\end{array}
\right)
\label{DetSol}
\end{align}
\smallskip\\
To prove this, it suffices to expand (\ref{DetSol}) by elements of the last two rows. For example, for $N = 3$ we have

\begin{align}
e_{3} =
\left|
\begin{array}{ccccccccccccccc}
e_1 & -g_1 & g_2 \\
\\
1 & c_1 & 1 \\
\\
0 & 1 & c_2 \\
\end{array}
\right| = c_2 \left|
\begin{array}{ccccccccccccccc}
e_1 & -g_1 \\
\\
1 & c_1 \\
\end{array}
\right| -  \left|
\begin{array}{ccccccccccccccc}
e_1 & g_2 \\
\\
1 & 1 \\
\end{array}
\right| = c_2 e_2 + g_2 - e_1
\end{align}
\smallskip\\
so that (\ref{CombinToda}) holds, for $N = 4$ we have

\begin{align}
e_{4} =
\left|
\begin{array}{ccccccccccccccc}
e_1 & -g_1 & g_2 & - g_3\\
\\
1 & c_1 & 1 & 0\\
\\
0 & 1 & c_2 & 1 \\
\\
0 & 0 & 1 & c_3 \\
\\
\end{array}
\right| = c_3 \left|
\begin{array}{ccccccccccccccc}
e_1 & -g_1 & g_2 \\
\\
1 & c_1 & 1 \\
\\
0 & 1 & c_2 \\
\end{array}
\right| - \left|
\begin{array}{ccccccccccccccc}
e_1 & -g_1 & -g_3 \\
\\
1 & c_1 & 0 \\
\\
0 & 1 & 1 \\
\end{array}
\right| = c_3 e_3 + g_3 - e_2
\end{align}
\smallskip\\
and (\ref{CombinToda}) holds again. Generalisation is obvious.

\subsection{Orthogonal polynomials}
If, instead, one expands the determinant by elements of the first row, one obtains

\begin{align}
e_{N} = e_1 \det
\left(
\begin{array}{ccccccccccccccc}
c_1 & 1 & 0 & \ldots \\
\\
1 & c_2 & 1 & \ldots \\
\\
0 & 1 & c_3 & \ldots \\
\\
\ldots & \ldots & \ldots & \ldots \\
\end{array}
\right) + \sum\limits_{i = 1}^{N-1} g_i \det
\left(
\begin{array}{ccccccccccccccc}
c_{i+1} & 1 & 0 & \ldots \\
\\
1 & c_{i+2} & 1 & \ldots \\
\\
0 & 1 & c_{i+3} & \ldots \\
\\
\ldots & \ldots & \ldots & \ldots \\
\end{array}
\right)
\label{DetSol2}
\end{align}
\smallskip\\
Therefore, the answer for the arbitrary $m$-point exponential correlation function can be written as

\begin{align}
e_N(x_1, \ldots, x_m) = T^{0}_{N}\Big((\Sigma x)^2\Big) \ e_1(x_1, \ldots, x_m) + \sum\limits_{i = 1}^{N-1} \ T^{i}_{N}\Big((\Sigma x)^2\Big) \ g_i(x_1, \ldots, x_m)
\label{GenSolOrt}
\end{align}
\smallskip\\
where $T^{i}_{j}(u)$ are special polynomials

\begin{align}
T^{i}_{j}(u) \ = \ \det\limits_{(j-1) \times (j-1)} \left(
\begin{array}{ccccccccccccccc}
2 + \dfrac{u}{i+1} & 1 & 0 & \ldots \\
\\
1 & 2 + \dfrac{u}{i+2} & 1 & \ldots \\
\\
0 & 1 & 2 + \dfrac{u}{i+3} & \ldots \\
\\
\ldots & \ldots & \ldots & \ldots \\
\end{array}
\right)
\label{Tijdet}
\end{align}
\smallskip\\
with a general formula

\begin{align}
T^i_j (u) = \sum\limits_{a = 0}^{(j-1)/2} \ \sum\limits_{k_1 + \ldots + k_{L} = 0}^{a} \ (-1)^a \prod\limits_{l = 1}^{L} \left( 2 + \dfrac{u}{i + l + 2 k_1 + \ldots + 2 k_l} \right), \ \ \ L = j - 2a - 1
\label{Tijexpl}
\end{align}
\smallskip\\
A few first polynomials $T^i_j(u)$ are:

\[
\begin{array}{ccc}
\\
T^0_0(u) = 0, \ \ \ T^1_0 = 0, \ \ \ T^2_0 = 0, \ \ \ \ldots\\
\\
T^0_1(u) = 1, \ \ \ T^1_1 = 1, \ \ \ T^2_1 = 1, \ \ \ \ldots\\
\\
T^0_2 = u + 2, \ \ \ T^1_2 = \dfrac{1}{2} u + 2, \ \ \ T^2_2 = \dfrac{1}{3} u + 2, \ \ \ \ldots\\
\\
T^0_3 = \dfrac{1}{2} u^2 + 3 u + 3, \ \ \ T^1_3 = \dfrac{1}{6} u^2 + \dfrac{5}{3} u + 3, \ \ \ T^2_3 = \dfrac{1}{12} u^2 + \dfrac{7}{6} u + 3, \ \ \ \ldots\\
\\
T^0_4 = \dfrac{1}{6} u^3 + 2 u^2 + 6 u + 4 , \ \ \ T^1_4 = \dfrac{1}{24} u^3 + \dfrac{3}{4} u^2 + \dfrac{43}{12} u + 4 , \ \ \ T^2_4 = \dfrac{1}{60} u^3 + \dfrac{2}{5} u^2 + \dfrac{13}{5} u + 4 , \ \ \ \ldots\\
\\
\end{array}
\]
It is easy to see, that polynomials $T^i_j(u)$ satisfy recursive relations

\begin{align}
\dfrac{u}{i+j} T^{i}_{j}(u) = T^{i}_{j+1}(u) - 2 T^{i}_{j}(u) + T^{i}_{j-1}(u)
\end{align}
\smallskip\\
which can be viewed as three-term relations in orthogonal polynomial theory. The three-term relation implies, that polynomials $T^i_j(u)$ form an orthogonal set of polynomials with respect to some, yet unidentified, local measure \cite{UFN3}. For example, for $i = 0$ they are orthogonal on the segment $(-\infty, 0)$ with measure $u e^{u} du$:

\begin{align} \int\limits_{-\infty}^{0} T^0_{j}(u) T^0_{k}(u) u e^{u} du = j \delta_{jk} \end{align}
\smallskip\\
i.e, for $i = 0$ they belong to the family of generalized Laguerre polynomials: $T^0_{N}(u) = {\rm Laguerre}^{(1)}_N \big( -u\big) $. It would be interesting to find the local measure, corresponding to polynomials $T^i_j(u)$ for arbitrary $i > 0$ (its existence is a consequence of three-term relations). In the next section, we derive an integral equation on it.

\subsection{The local measure}

For convenience, let us introduce normalised (with unit leading coefficient) polynomials

\begin{align}Q^i_j(u) = \dfrac{(i+j-1)!}{i!} T^i_j(u) = u^{j-1} + \ldots\end{align}
\smallskip\\
which satisfy recursive relations

\begin{align}
u Q^{i}_{j}(u) = Q^{i}_{j+1}(u) - 2 (i + j) Q^{i}_{j}(u) + \dfrac{i+j}{i+j-2} Q^{i}_{j-1}(u)
\end{align}
\smallskip\\
A first few polynomials $Q^i_j(u)$ are
\[
\begin{array}{ccc}
\\
Q^i_0(u) = 0\\
\\
Q^i_1(u) = 1\\
\\
Q^i_2(u) = u+2 i+2\\
\\
Q^i_3(u) = u^2+(4 i+6) u+3 i^2+9 i+6\\
\\
\end{array}
\]
and so on. For each particular value of $i$, the system of polynomials $\big\{Q^i_j\big\}, j = 0,1,2,\ldots$ is orthogonal with respect to the unknown measure $d\mu_i(u) = \omega_i(u)du$. The orthogonality relations can be written as

$$ \int\limits_{-\infty}^{0} Q^i_j(u) Q^i_k(u) \omega_i(u) du = 0, \ \ \ \ j \neq k$$
\smallskip\\
(we assume that the segment is always $(-\infty, 0)$, just like for $i = 0$). The moments $M_k(i) = \int\limits_{-\infty}^{0} u^k \omega_i(u) du$ can be found from these orthogonality relations. Several first moments, obtained in this way, are

\begin{align}
M_0(i) = 1, \ \ M_1(i) = -(2i + 2), \ \ M_2(i) = 5 i^2+11 i+6, \ \ M_3(i) = -(14 i^3+50 i^2+60 i+24), \ \ \ldots
\end{align}
\smallskip\\
Direct calculation shows, that generating function for these moments has a form

\begin{align}
\begin{array}{ccc}
\sum\limits_{k = 0}^{\infty} (-1)^k M_k \dfrac{z^{k+1}}{k+1} \ = \ z + (2i + 2) \dfrac{z^2}{2} + (5 i^2+11 i+6) \dfrac{z^3}{3} + (14 i^3+50 i^2+60 i+24) \dfrac{z^4}{4} + \ldots = \\
\\
= \dfrac{1}{i(i+1)} \log\left( 1 + i(i+1) z + \dfrac{i(i+1)^2(i+2)}{2} z^2 + \dfrac{i(i+1)^2(i+2)^2(i+3)}{6} z^3 + \ldots \right)\\
\end{array}
\end{align}
\smallskip\\
This relation can be written as an integral equation

\begin{align} \int\limits_{-\infty}^{0} \log (1 + uz) \dfrac{\omega_i(u)}{u} du = \dfrac{1}{i(i+1)} \log\left( \sum\limits_{k = 0}^{\infty} \dfrac{(k+i)!(k+i-1)!}{i!(i-1)!} \dfrac{z^{k}}{k!} \right) = \dfrac{1}{i(i+1)} \log \ _{2}F_{0}\left( i, i+1; z \right) \end{align}
\smallskip\\
where $_{2}F_{0}$ is the generalized hypergeometric function. The local measure $\omega_i(u)du$ can be found as its solution. It in present paper, we do not solve this equation and we do not use this measure in what follows.

\subsection{1,2,3-point functions for particular N}

Application of eq. (\ref{GenSolOrt}) to particular correlation functions is straightforward. For $m = 1$ we have

\begin{align}
 e_N(x) = T^{0}_{N}\Big(x^2\Big) e_1(x)
\label{Expo1p}
\end{align}
\smallskip\\
for $m = 2$ we have

\begin{align}
 e_N(x,y) = T^{0}_{N}\Big((x+y)^2\Big) e_1(x,y) - \sum\limits_{i = 1}^{N-1} \dfrac{(xy)^2}{i^2} T^{i}_{N}\Big((x+y)^2\Big) e_i(x) e_i(y)
\label{Expo2p}
\end{align}
\smallskip\\
for $m = 3$ we have

\begin{align} e_N(x,y,z) \ = \ T^{0}_{N}\Big((x+y+z)^2\Big) e_1(x,y,z) + 2 \sum\limits_{i = 1}^{N-1} T^{i}_{N}\Big((x+y+z)^2\Big) \dfrac{(xyz)^2}{i^3} e_i(x) e_i(y) e_i(z) - \label{Expo3p}\end{align}

$$
 - \sum\limits_{i = 1}^{N-1} T^{i}_{N}\Big((x+y+z)^2\Big) \left( \dfrac{(xz + yz)^2}{i^2} e_i(x + y) e_i(z) + \dfrac{(xy + yz)^2}{i^2} e_i(x + z) e_i(y) +\dfrac{(xy + xz)^2}{i^2} e_i(y + z) e_i(x) \right)
$$
\smallskip\\
and so on. In this way, all the exponential correlation functions $e_N(x_1, \ldots, x_m)$ are expressed through polynomials $T^i_j(u)$. These expressions are fully explicit and constructive (since all kinds of explicit formulas are available for $T^i_j(u)$) but they are not very illuminating, certainly less beautiful than eq.(\ref{HarerZagierDuo}).

\subsection{Universal 1,2,3-point functions}
To cure this problem, one may switch to universal $\lambda$-dependent correlation functions, hoping that this will improve the situation. After conversion to universal functions

\begin{align}e(\lambda; x_1, \ldots, x_m) = \sum\limits_{N = 0}^{\infty} \ \lambda^N \ e_{N}(x_1, \ldots, x_m)\end{align}
\smallskip\\
the Toda equation (\ref{CombinToda}) transforms into a differential equation of first order:

\begin{align}
 \lambda \dfrac{\partial}{\partial \lambda} \left( \dfrac{(1 - \lambda)^2}{\lambda} e(\lambda; x_1, \ldots, x_m) - g(\lambda; x_1, \ldots, x_m) \right) = \big(\Sigma x\big)^2 e(\lambda; x_1, \ldots, x_m)
\label{DifferentTodaE}
\end{align}
\smallskip\\
where $\Sigma x = x_1 + \ldots + x_m$ and $g(\lambda; x_1, \ldots, x_m)$ is the differential equation's free term:

\begin{align}g(\lambda; x_1, \ldots, x_m) = \sum\limits_{N = 0}^{\infty} \ \lambda^N \ g_{N}(x_1, \ldots, x_m)\end{align}
\smallskip\\
Unique solution of eq. (\ref{DifferentTodaE}), satisfying the usual $N = 1$ initial condition

\begin{align}
\dfrac{\partial}{\partial \lambda} e(\lambda; x_1, \ldots, x_m)\Big|_{\lambda = 0} = e_1(x_1, \ldots, x_m)
\end{align}
is given by

\begin{align}
e(\lambda; {\vec x}) = \dfrac{\lambda}{(1-\lambda)^2} \exp\left( \dfrac{(\Sigma x)^2 \lambda}{1 - \lambda} \right) e_1({\vec x}) + \dfrac{\lambda}{(1-\lambda)^2} \int\limits_{0}^{\lambda} d t \ \dfrac{\partial g(t; {\vec x})}{\partial t} \ \exp\left( \dfrac{(\Sigma x)^2 (\lambda - t) }{(1 - \lambda)(1-t)} \right)
\end{align}
\smallskip\\
Application of this formula to particular correlation functions is straightforward. Let us begin with the simplest case, i.e, with the 1-point correlation function. Since $g(\lambda; x) = 0$, it immediately follows, that

$$
 e(\lambda, x) = \dfrac{\lambda}{(1-\lambda)^2} \ \exp\left( \dfrac{x^2 \lambda}{1 - \lambda} \right) e_1(x)
$$
\smallskip\\
Moreover, the initial function can be calculated explicitly:

\begin{align}
 e_1(x) = \int\limits_{-\infty}^{+\infty} d\phi e^{-\phi^2/2 + x \phi } = e^{x^2/2}
\end{align}
so that

\begin{equation}
\addtolength{\fboxsep}{5pt}
\boxed{
\begin{gathered}
 e(\lambda, x) = \sum\limits_{N = 0}^{\infty} \ \lambda^N \ e_N(x) = \dfrac{\lambda}{(1-\lambda)^2} \ \exp\left( \dfrac{x^2}{2} \cdot \dfrac{1 + \lambda}{1 - \lambda} \right)
\end{gathered}
}\label{Expo1p}
\end{equation}
\smallskip\\
This answer is consistent with the Harer-Zagier 1-point function (\ref{HarerZagierUni}) and transformation (\ref{ContourTransform}):

\begin{align*} e(\lambda; x) = \oint \dfrac{(1 + y) dy}{y} \ \varphi(\lambda; y) \exp\left( \dfrac{x^2}{2y^2} \right) = \emph{} \end{align*}

\begin{align}
\emph{} = \dfrac{\lambda}{(1 - \lambda)^2} \oint \dfrac{(1 + y) dy}{y} \ \dfrac{ \exp\left( \dfrac{x^2}{2y^2} \right) }{\dfrac{1 - \lambda}{1 + \lambda} - y^2 } = \dfrac{\lambda}{(1 - \lambda)^2} \oint \dfrac{dy}{y} \ \dfrac{ \exp\left( \dfrac{x^2}{2y^2} \right) }{\dfrac{1 - \lambda}{1 + \lambda} - y^2 } = \dfrac{\lambda}{(1 - \lambda)^2} \exp\left( \dfrac{x^2}{2} \cdot \dfrac{1 + \lambda}{1 - \lambda} \right)
\end{align}
\smallskip\\
The last equality is a direct corollary of Cauchy residue theorem. Let us turn to the next-to-simplest case, i.e, to the exponential 2-point function. In this case, the free term $g(\lambda; x,y)$ is no longer zero:

\begin{align} \nonumber g(\lambda; x,y) \ = \ & - \sum\limits_{N = 0}^{\infty} \ \lambda^N \ \dfrac{x^2y^2}{N^2} e_N(x) e_N(y) = \\ \noalign{\medskip} \nonumber & \emph{} = - \oint\oint \dfrac{du_1 du_2}{u_1 u_2} \left( \sum\limits_{N = 0}^{\infty} u_1^{-N} u_2^{-N} \lambda^N \right) \exp\left( \dfrac{x^2}{2} \cdot \dfrac{1 + u_1}{1 - u_1} + \dfrac{y^2}{2} \cdot \dfrac{1 + u_2}{1 - u_2} \right) = \\ \noalign{\medskip}  & \emph{} = \oint\oint \dfrac{du_1 du_2}{\lambda - u_1 u_2} \exp\left( \dfrac{x^2}{2} \cdot \dfrac{1 + u_1}{1 - u_1} + \dfrac{y^2}{2} \cdot \dfrac{1 + u_2}{1 - u_2} \right) \end{align}
\smallskip\\
Consequently, the 2-point function can be written as

\begin{equation}
\addtolength{\fboxsep}{5pt}
\boxed{
\begin{gathered}
\dfrac{(1-\lambda)^2}{\lambda} \ e(\lambda; x,y) = \exp\left( \dfrac{(x + y)^2 \lambda}{1 - \lambda} \right) e_1(x,y) - \emph{}\\
\\
\emph{} - \int\limits_{0}^{\lambda} d t \oint\oint \dfrac{du_1 du_2}{(t - u_1 u_2)^2} \exp\left( \dfrac{(x+y)^2 (\lambda - t) }{(1 - \lambda)(1-t)} + \dfrac{x^2}{2} \cdot \dfrac{1 + u_1}{1 - u_1}  + \dfrac{y^2}{2} \cdot \dfrac{1 + u_2}{1 - u_2}  \right)
\end{gathered}
}\label{Expo2p}
\end{equation}
\smallskip\\
where

$$e_1(x,y) = \int\limits_{-\infty}^{+\infty} d\phi e^{-\phi^2/2 + x \phi + y \phi} - \left( \int\limits_{-\infty}^{+\infty} d\phi e^{-\phi^2/2 + x \phi } \right) \left( \int\limits_{-\infty}^{+\infty} d\phi e^{-\phi^2/2 + y \phi } \right) = e^{(x+y)^2/2} - e^{x^2/2} e^{y^2/2}$$
\smallskip\\
In principle, it should be possible to obtain the same formula with two contour integrals in a different way, directly applying the transformation (\ref{ContourTransform}) to the Harer-Zagier 2-point function. Note, that (\ref{Expo2p}) is far less concise, than (\ref{HarerZagierDuo}). Instead, its advantage is the possibility of generalization: it is not a problem to write its analogue for any $m$-point function. Say, for the 3-point function we have

\begin{equation}
\addtolength{\fboxsep}{5pt}
\boxed{
\begin{gathered}
\dfrac{(1-\lambda)^2}{\lambda} \ e(\lambda; x,y,z) = \exp\left( \dfrac{(x + y + z)^2 \lambda}{1 - \lambda} \right) e_1(x,y,z) + \emph{}\\
\\
\emph{} + \int\limits_{0}^{\lambda} d t \oint\oint \dfrac{(y+z)^2 du_1 du_2}{t(t - u_1 u_2)} \exp\left( \dfrac{(x+y+z)^2 (\lambda - t) }{(1 - \lambda)(1-t)} + \dfrac{x^2}{2} \cdot \dfrac{1 + u_1}{1 - u_1} \right)  e(u_2; y,z) + \emph{}\\
\\
\emph{} + \int\limits_{0}^{\lambda} d t \oint\oint \dfrac{(x+z)^2 du_1 du_2}{t(t - u_1 u_2)} \exp\left( \dfrac{(x+y+z)^2 (\lambda - t) }{(1 - \lambda)(1-t)} + \dfrac{y^2}{2} \cdot \dfrac{1 + u_1}{1 - u_1} \right)  e(u_2; x,z) + \emph{}\\
\\
\emph{} + \int\limits_{0}^{\lambda} d t \oint\oint \dfrac{(x+y)^2 du_1 du_2}{t(t - u_1 u_2)} \exp\left( \dfrac{(x+y+z)^2 (\lambda - t) }{(1 - \lambda)(1-t)} + \dfrac{z^2}{2} \cdot \dfrac{1 + u_1}{1 - u_1} \right)  e(u_2; x,y) + \emph{}\\
\\
+ \int\limits_{0}^{\lambda} d t \oint\oint\oint \dfrac{2 du_1 du_2 du_3}{(t - u_1 u_2 u_3)^2} \exp\left( \dfrac{(x+y+z)^2 (\lambda - t) }{(1 - \lambda)(1-t)} + \dfrac{x^2}{2} \dfrac{1 + u_1}{1 - u_1} + \dfrac{y^2}{2} \dfrac{1 + u_2}{1 - u_2} + \dfrac{z^2}{2} \dfrac{1 + u_3}{1 - u_3} \right)
\end{gathered}
}\label{Expo3p}
\end{equation}
\smallskip\\
where

$$e_1(x,y,z) = e^{(x+y+z)^2/2} - e^{(x+y)^2/2} e^{z^2/2} - e^{(x+z)^2/2} e^{y^2/2} - e^{(y+z)^2/2} e^{x^2/2} + 2 e^{x^2/2} e^{y^2/2} e^{z^2/2}$$
\smallskip\\
Clearly, functions (\ref{Expo1p}), (\ref{Expo2p}) and (\ref{Expo3p}) belong to a family of exact solutions, which are less elegant, than (\ref{HarerZagierDuo}).

\section{Standard correlation functions (resolvents) }

\subsection{Genus expansion}

Among the all correlation functions, the most widely used ones are the standard correlation functions:

\begin{align} \rho_{N}(x_1, \ldots, x_m) = \Big< \Big< \tr \left( \dfrac{1}{1 - x_1 \phi} \right) \ldots \tr \left( \dfrac{1}{1 - x_m \phi} \right) \Big> \Big> = \sum\limits_{i_1 \ldots i_m = 0}^{\infty} K_{i_1 \ldots i_m}(N) \ x_1^{i_1} \ldots x_m^{i_m} \end{align}
\smallskip\\
As we already mentioned in s.3, they are better known as resolvents and usually defined as

\begin{align} W_{N}(x_1, \ldots, x_m) = \Big< \Big< \tr \left( \dfrac{1}{x_1 - \phi} \right) \ldots \tr \left( \dfrac{1}{x_m - \phi} \right) \Big> \Big> = \sum\limits_{i_1 \ldots i_m = 0}^{\infty} K_{i_1 \ldots i_m}(N) \ x_1^{-i_1-1} \ldots x_m^{-i_m-1} \end{align}
\smallskip\\
All-genera resolvents are divergent series, because the number of fat graphs of arbitrary genus grows rapidly. This property is typical for any  perturbation theory, where the number of Feynman diagrams of order $n$ usually grows as $n!$. In practice this means that one can at best hope to represent the full $\rho_{N}$ as an integral, just like it happens with the archetypical divergent sum

$$ \sum\limits_{n = 0}^{\infty} n! x^n = \int\limits_{0}^{\infty} \dfrac{e^{-t} dt}{1-tx} $$
\smallskip\\
which is actually divergent only for $x \in R_{+}$. Such integral representation for resolvents is naturally provided by Harer-Zagier correlation functions: if the latter are known, the resolvents are given by

\begin{align} \rho_{N}(x_1, \ldots, x_m) = \int\limits_{-\infty}^{\infty} (1 + y_1) dy_1 \ldots \int\limits_{-\infty}^{\infty} (1 + y_m) dy_m \ \varphi_{N}(x_1 y_1, \ldots, x_m y_m) \ e^{- y_1^2/2 - \ldots - y_m^2/2} \label{h24} \end{align}
\smallskip\\
Another way to deal with divergence of resolvents, often used in practice, is to introduce the genus expansion: namely, to consider generating functions for fat graphs of fixed genus $g$:

\begin{align} \rho_{(g)}(x_1, \ldots, x_m) = \sum\limits_{i_1 \ldots i_m = 0}^{\infty} K^{(g)}_{i_1 \ldots i_m} \ x_1^{i_1} \ldots x_m^{i_m} \end{align}
\smallskip\\
where

\begin{align}
K^{(g)}_{i_1 \ldots i_m} = \mbox{ coefficient of } N^{\deg(g)} \mbox{ in } K_{i_1 \ldots i_m}(N), \ \ \ \deg(g) = (i_1 + \ldots + i_m)/2 + (2 - 2g) - m
\label{PowerCounting}
\end{align}
\smallskip\\
Recall, that genus $g$ contribution to the connected correlator of $\tr \phi^{i_1} \ldots \tr \phi^{i_m}$ scales as $N$ to the power $(i_1 + \ldots + i_m)/2 + (2 - 2g) - m$. Defined in this way, the genus $g$ standard correlation functions (the genus $g$ resolvents) are no longer divergent: the number of fat graphs of fixed genus grows much slower, than the total number of fat graphs. We will now demonstrate that relation (\ref{h24}) indeed allows to describe the genus expansion and find resolvents for any particular genus $g$.

\subsection{1-point function}

The exact 1-point resolvent is given by

\begin{align} \rho_{N}(x) = \int\limits_{-\infty}^{\infty} dy \ e^{- y^2/2} \  \varphi_{N}(xy) \label{Hz1} \end{align}
\smallskip\\
where the 1-point Harer-Zagier function is given by

\begin{align}
\varphi_N(x) = \dfrac{1}{2x^2} \left( \left(\dfrac{1 + x^2}{1 - x^2}\right)^N - 1 \right)
\end{align}
\smallskip\\
Technically, to extract the genus expansion it is most convenient to introduce another variable $X = x \sqrt{N}$ (see s. IV of \cite{AMM}). As a consequence of the scaling rule (\ref{PowerCounting}), in terms of $X$ the genus expansion becomes simply the $1/N$ expansion. Indeed, in terms of $X$ the Harer-Zagier function takes form

$$ \varphi_N\left( \dfrac{X}{\sqrt{N}} \right) = \dfrac{N}{2X^2} \left( \left(\dfrac{N + X^2}{N - X^2}\right)^N - 1 \right) $$
\smallskip\\
and posesses an expansion in negative powers of $N$:

$$\left(\dfrac{N + X^2}{N - X^2}\right)^N = \exp \left\{ N \log \left( \dfrac{N + X^2}{N - X^2} \right) \right\} = \exp \left( 2X^2 + \dfrac{2X^6}{3N^2} + \dfrac{2X^{10}}{5N^4} + \ldots \right) = \exp \left( \sum\limits_{k = 0}^{\infty} \dfrac{2X^{4k+2}}{(2k+1) N^{2k}} \right) $$
\smallskip\\
The exponent of a series can be represented as a series again:

$$\exp \left( 2X^2 + \dfrac{2X^6}{3N^2} + \dfrac{2X^{10}}{5N^4} + \ldots \right) = e^{2X^2} \cdot \left( 1 + \dfrac{2X^6}{3N^2} + \dfrac{\dfrac{2}{5} X^{10} + \dfrac{2}{9} X^{12}}{N^4} + \dfrac{\dfrac{2}{7}X^{14} + \dfrac{4}{15} X^{16} + \dfrac{4}{81} X^{18} }{N^6} + \ldots \right)$$
\smallskip\\
or, in a general form,

$$ \exp \left( \sum\limits_{k = 0}^{\infty} \dfrac{2X^{4k+2}}{(2k+1) N^{2k}} \right) = e^{2X^2} \cdot \sum\limits_{p = 0}^{\infty} \sum\limits_{q = 0}^{p} \dfrac{1}{q!} \dfrac{X^{4p+2q}}{N^{2p}} \sum\limits_{i_1 + \ldots + i_q = p} \dfrac{2^q}{(2i_1+1) \ldots (2i_q + 1)}$$
\smallskip\\
Accordingly, the Harer-Zagier function takes form

$$
\varphi_N\left( \dfrac{X}{\sqrt{N}} \right) = N \dfrac{e^{2X^2} - 1}{2X^2} + \dfrac{e^{2X^2}}{N} \dfrac{X^4}{3} + \dfrac{e^{2X^2}}{N^3} \left( \dfrac{X^8}{5} + \dfrac{X^{10}}{9} \right) + \dfrac{e^{2X^2}}{N^6} \left( \dfrac{X^{12}}{7} + \dfrac{2X^{14}}{15} + \dfrac{2X^{16}}{81} \right) + \ldots
$$
\smallskip\\
or, generally,

\begin{align}
\varphi_N\left( \dfrac{X}{\sqrt{N}} \right) = N \dfrac{e^{2X^2} - 1}{2X^2} + e^{2X^2} \cdot \sum\limits_{p = 1}^{\infty} \sum\limits_{q = 1}^{p} \dfrac{1}{q!} \dfrac{X^{4p+2q-2}}{N^{2p-1}} \sum\limits_{i_1 + \ldots + i_q = p} \dfrac{2^{q-1}}{(2i_1+1) \ldots (2i_q + 1)}
\end{align}
\smallskip\\
The leading contribution here is the genus 0 Harer-Zagier correlation function:

\begin{align}\varphi_{(0)}(X) = \dfrac{e^{2X^2} - 1}{2X^2}\end{align}
\smallskip\\
Accordingly, the next-to-leading contribution corresponds to genus 1, and so on:

\begin{align}
\varphi_{(1)}(X) \ = \ & \dfrac{X^4}{3} e^{2X^2}\\
\nonumber\\
\varphi_{(2)}(X) \ = \ & \left( \dfrac{X^8}{5} + \dfrac{X^{10}}{9} \right) e^{2X^2}\\
\nonumber\\
\varphi_{(3)}(X) \ = \ & \left( \dfrac{X^{12}}{7} + \dfrac{2X^{14}}{15} + \dfrac{2X^{16}}{81} \right) e^{2X^2} \\
\nonumber\\
\nonumber \ldots\\
\nonumber\\
\varphi_{(g)}(X) \ = \ & \sum\limits_{q = 1}^{g} \dfrac{X^{4g+2q-2} e^{2X^2}}{q!} \sum\limits_{i_1 + \ldots + i_q = g} \dfrac{2^{q-1}}{(2i_1+1) \ldots (2i_q + 1)}
\end{align}
\smallskip\\
Passing back to resolvents by taking Gaussian integrals (\ref{Hz1}), we obtain in genus zero

\begin{align}
\rho_{(0)}(X) = \int\limits_{-\infty}^{\infty} dY \ e^{- Y^2/2} \ \dfrac{e^{2X^2Y^2} - 1}{2X^2Y^2} = \dfrac{1 - \sqrt{1 - 4X^2}}{2X^2}
\end{align}
\smallskip\\
This is the celebrated Wigner semi-circle distribution \cite{Wigner}. Similarly, in higher genera we obtain

\begin{align}
\rho_{(1)}(X) \ = \ & 3 X^4 \big(1 - 4 X^2\big)^{-5/2}\\
\nonumber \\
\rho_{(2)}(X) \ = \ & \big( 21 X^8 + 21 X^{10} \big) \big(1 - 4 X^2\big)^{-11/2} \\
\nonumber \\
\rho_{(3)}(X) \ = \ & \big( 1485 X^{12} + 6138 X^{14} + 1738 X^{16} \big) \big(1 - 4 X^2\big)^{-17/2} \\
\nonumber \\
\nonumber \ldots\\
\nonumber \\
\rho_{(g)}(X) \ = \ & \sum\limits_{q = 1}^{g} \dfrac{(4g + 2q - 3)!! X^{4g+2q-2} (1 - 4X^2)^{1/2 - 2g - q}}{q!} \sum\limits_{i_1 + \ldots + i_q = g} \dfrac{2^{q-1}}{(2i_1+1) \ldots (2i_q + 1)} \\
\nonumber
\end{align}
As one can see, it is a straightforward exercise to extract the genus expansion from the Harer-Zagier 1-point function -- it is even possible to write a formula for arbitrary $g$. Using the main result of present paper -- the exact 2-point Harer-Zagier function -- we can do a similar calculation at the 2-point level.

\subsection{2-point function}

The exact 2-point odd and even resolvents are given by

\begin{align} \rho^{+}_{N}(x,y) = \int\limits_{-\infty}^{\infty} du \int\limits_{-\infty}^{\infty} dv \ \exp\left( - \dfrac{u^2 + v^2}{2} \right) \ \varphi^{+}_{N}(x u, y v)
\label{Hx21}
\end{align}

\begin{align} \rho^{-}_{N}(x,y) = \int\limits_{-\infty}^{\infty} u du \int\limits_{-\infty}^{\infty} v dv \ \exp\left( - \dfrac{u^2 + v^2}{2} \right) \ \varphi^{-}_{N}(x u, y v)
\label{Hx22}
\end{align}
\medskip\\
where the 2-point odd and even Harer-Zagier functions are given by (\ref{U1}) and (\ref{U2}):

\begin{align*}
 \varphi^{-}_N(x,y) = \int\limits_{0}^{xy} \dfrac{dt}{2(x^2+y^2)} \ \left( \left( \dfrac{1 + t^2 + x^2 + y^2}{1+t^2-x^2-y^2} \right)^N - 1 \right)
\end{align*}

\begin{align*}
\varphi^{+}_N(x,y) = \int\limits_{0}^{xy} \dfrac{xy dt}{(x^2+y^2)^2} \ \left( \left( \dfrac{1 + t^2 + x^2 + y^2}{1+t^2-x^2-y^2} \right)^N \dfrac{2N(x^2+y^2)(1+t^2)+(x^2+y^2)^2-(1+t^2)^2}{(1 + t^2 + x^2 + y^2)(1 + t^2 - x^2 - y^2)} + 1 \right)
\end{align*}
\medskip\\
The full resolvent is a sum of odd and even parts:

\begin{align} \rho_{N}(x,y) = \rho^{(+)}_{N}(x,y) + \rho^{(-)}_{N}(x,y) \end{align}
\medskip\\
Again, to extract the genus expansion we introduce another variables $X = x \sqrt{N}, Y = y \sqrt{N} $. As a consequence of the scaling rule (\ref{PowerCounting}), in terms of $X$ and $Y$ the genus expansion becomes simply the $1/N$ expansion. Taking the intermediate $t$-integral (we do not go into details here) we obtain in the odd case

\begin{align*}
 \varphi^{-}_N\left(\dfrac{X}{\sqrt{N}},\dfrac{Y}{\sqrt{N}}\right) \ = \ & \dfrac{XY \Big(e^{2X^2 + 2Y^2} - 1\Big)}{2(X^2+Y^2)} + \dfrac{e^{2X^2 + 2Y^2}}{N^2} \left( \dfrac{X^3 Y^3}{3} + \dfrac{X^5 Y}{3} + \dfrac{X Y^5}{3} \right) + \ldots
\end{align*}
\medskip\\
and in the even case

\begin{align*}
 \varphi^{+}_N\left(\dfrac{X}{\sqrt{N}},\dfrac{Y}{\sqrt{N}}\right) \ = \ & \dfrac{X^2Y^2}{(X^2+Y^2)^2} \left((2X^2 + 2Y^2 - 1) e^{2X^2 + 2Y^2} - 1\right) + \emph{} \\ & \\ & \emph{} + \dfrac{e^{2X^2 + 2Y^2}}{N^2} \left( \dfrac{4}{3} X^6 Y^2 + \dfrac{4}{3} X^4 Y^4 +\dfrac{4}{3} X^2 Y^6 +\dfrac{4}{3} X^4 Y^2 +\dfrac{4}{3} X^2 Y^4 \right) + \ldots
\end{align*}
\medskip\\
We stop at genus one and do not write the genus two and higher contributions, it is absolutely straightforward to obtain them as higher $1/N$ corrections. To write a formula for arbitrary $g$ (like we did in the 1-point case) is a more involved, but still feasible exercise, which remains to be done. For the Harer-Zagier functions in genus zero, we obtain

\begin{align}\varphi^{+}_{(0)}(X,Y) = \dfrac{X^2Y^2}{(X^2+Y^2)^2} \left((2X^2 + 2Y^2 - 1) e^{2X^2 + 2Y^2} - 1\right)\end{align}

\begin{align}\varphi^{-}_{(0)}(X,Y) = \dfrac{XY \Big(e^{2X^2 + 2Y^2} - 1\Big)}{2(X^2+Y^2)}\end{align}
\smallskip\\
In genus one we have

\begin{align}\varphi^{+}_{(1)}(X,Y) = \left( \dfrac{4}{3} X^6 Y^2 + \dfrac{4}{3} X^4 Y^4 +\dfrac{4}{3} X^2 Y^6 +\dfrac{4}{3} X^4 Y^2 +\dfrac{4}{3} X^2 Y^4 \right) e^{2X^2 + 2Y^2}\end{align}

\begin{align}\varphi^{-}_{(1)}(X,Y) = \left( \dfrac{X^3 Y^3}{3} + \dfrac{X^5 Y}{3} + \dfrac{X Y^5}{3} \right) e^{2X^2 + 2Y^2} \end{align}
\smallskip\\
Passing back to resolvents by taking Gaussian integrals (\ref{Hx21}) and (\ref{Hx22}), we obtain in genus zero

\begin{align}\rho^{+}_{(0)}(X,Y) = \dfrac{X^2Y^2}{(X^2-Y^2)^2} \left( \dfrac{1 - 2X^2 - 2Y^2}{\sqrt{1-4X^2}\sqrt{1-4Y^2}} - 1 \right) \end{align}

\begin{align}\rho^{-}_{(0)}(X,Y) = \dfrac{XY}{2(X^2-Y^2)^2} \left( \dfrac{X^2 + Y^2 - 8X^2Y^2}{\sqrt{1-4X^2}\sqrt{1-4Y^2}} - X^2 - Y^2 \right) \end{align}
\smallskip\\
and in genus one

\begin{align}\rho^{+}_{(1)}(X,Y) = \dfrac{8 X^2 Y^4+8 X^4 Y^2+8 X^2 Y^6-104 X^4 Y^4+8 X^6 Y^2-32 X^4 Y^6-32 X^6 Y^4+640 X^6 Y^6}{(1-4X^2)^{7/2} (1-4Y^2)^{7/2}} \end{align}

\begin{align}\rho^{-}_{(1)}(X,Y) = \dfrac{5 X Y^5+3 X^3 Y^3+5 X^5 Y-52 X^3 Y^5-52 X^5 Y^3+208 X^5 Y^5}{(1-4X^2)^{7/2} (1-4Y^2)^{7/2}}\end{align}
\smallskip\\
Similarly, contribution of any higher genus can be found by expanding the exact functions (\ref{U1}) and (\ref{U2}). This method should be also applicable to 3-point and higher resolvents, but, unfortunately, 3-point and higher analogues of (\ref{U1}) and (\ref{U2}) are not found yet.

\section{Conclusion}

Despite the apparent simplicity and transparency of the Gaussian Hermitian model, which is beyond any doubt one of the most studied and best understood matrix models, its correlators form a complicated combinatorial system. Given a family of correlators, we can rarely explicitly describe its behaviour. Miriads of integer numbers, counting appropriate fat graphs or discrete Riemann surfaces, appear in a seemingly random fashion. Integer numbers form patterns, they grow and they change according to laws which, despite the model is Gaussian, are far from being simple. In the case of one-point correlators

\[
\begin{array}{ccccc}
\\
\Big<\Big< \tr \phi^2 \Big> \Big> = N^2\\
\\
\Big<\Big< \tr \phi^4 \Big> \Big> = 2N^3 + N\\
\\
\Big<\Big< \tr \phi^6 \Big> \Big> = 5N^4 + 10N^2\\
\\
\Big<\Big< \tr \phi^8 \Big> \Big> = 14 N^5+70 N^3+21 N\\
\\
\Big<\Big< \tr \phi^{10} \Big> \Big> = 42 N^6+420 N^4+483 N^2\\
\\
\ldots \\
\end{array}
\]
\smallskip\\
these laws can be summarised in one compact formula, found by Harer and Zagier:

$$ \dfrac{ \Big<\Big< \tr \phi^{2k} \Big> \Big> }{(2k-1)!!} = \mbox{ coefficient of } x^{2k} \lambda^N \mbox{ in } \dfrac{\lambda}{1 - \lambda} \dfrac{1}{(1 - \lambda) - (1 + \lambda) x^2} $$
\smallskip\\
The modest aim of our research was to find analogous formula for the two-point correlators

\[
\begin{array}{ccccc}
\Big<\Big< \tr \phi \ \tr \phi \Big> \Big> = N \ \ \ \ \Big<\Big< \tr \phi \ \tr \phi^3 \Big> \Big> = 3 N^2 \ \ \ \ \Big<\Big< \tr \phi \ \tr \phi^5 \Big> \Big> = 10 N^3+5 N \\ \\
\Big<\Big< \tr \phi \tr \phi^7 \Big> \Big> = 35 N^4+70 N^2 \\ \\ \Big<\Big< \tr \phi^3 \ \tr \phi^3 \Big> \Big> = 12 N^3+3 N \ \ \ \ \Big<\Big< \tr \phi^3 \ \tr \phi^5 \Big> \Big> = 45 N^4+60 N^2  \\ \\ \Big<\Big< \tr \phi^3 \ \tr \phi^7 \Big> \Big> = 168 N^5+630 N^3+147 N \\
\\
\Big<\Big< \tr \phi^5 \ \tr \phi^5 \Big> \Big> = 180 N^5+600 N^3+165 N \ \ \ \ \Big<\Big< \tr \phi^5 \ \tr \phi^7 \Big> \Big> = 700 N^6+4900 N^4+4795 N^2 \\
\\
\Big<\Big< \tr \phi^7 \ \tr \phi^7 \Big> \Big> = 2800 N^7+34300 N^5+81340 N^3+16695 N\\
\\
\ldots
\end{array}
\]
\smallskip\\
and it appears to be

$$ \dfrac{ \Big<\Big< \tr \phi^{2i+1} \tr \phi^{2j+1}  \Big> \Big> }{(2i+1)!!(2j+1)!!} = \mbox{ coefficient of } x^{2k+1} y^{2m+1} \lambda^N \mbox{ in } \dfrac{\lambda}{(\lambda-1)^{3/2}} \dfrac{\arctan\left( \dfrac{xy \sqrt{\lambda-1}}{\sqrt{\lambda - 1 + (\lambda + 1) (x^2 + y^2)}} \right)}{\sqrt{\lambda - 1 + (\lambda + 1) (x^2 + y^2)}} $$
\smallskip\\
This is of course just the first step (or, better to say, the second step). Three-point and higher correlators are still under-investigated. We are yet very far from complete understanding of integer numbers related to fatgraphs: hopefully, many more compact and beautiful formulas lie in wait.

\section{Acknowledgements}

We are grateful to D.Vasiliev for stimulating discussions. Our work is partly supported by Russian Federal Nuclear Energy Agency
and the Russian President's Grant of Support for the Scientific Schools NSh-3035.2008.2, by RFBR grant 07-02-00547,
by the joint grants 09-01-92440-CE, 09-02-91005-ANF and 09-02-93105-CNRS. The work of Sh.Shakirov is also supported in part by the Moebius Contest Foundation for Young Scientists and by the Dynasty Foundation.

\end{document}